\documentclass[conference]{IEEEtran}

\IEEEoverridecommandlockouts
\usepackage{cite}
\usepackage{amsmath,amssymb,amsfonts}
\usepackage{algorithmic}
\usepackage{graphicx}
\usepackage{textcomp}
\usepackage{xcolor}
\def\BibTeX{{\rm B\kern-.05em{\sc i\kern-.025em b}\kern-.08em
    T\kern-.1667em\lower.7ex\hbox{E}\kern-.125emX}}

\usepackage{listings}
\lstdefinelanguage{sial}
   {morekeywords = {sial,endsial,proc,endproc,return,call,pardo,endpardo,also,
    do,enddo,cycle,exit,if,else,endif,put,get,prepare,request,prequest,collective,
    execute,aoindex,moindex,moaindex,mobindex,	index,laindex,scalar,static,temp,
    local,distributed,served,create,delete,allocate,deallocate,compute_integrals,
    destroy,where,subindex,of,in, sip_barrier, predefined, special, collective_sum,
    assert_same, contiguous, broadcast_static
    gpu_begin, gpu_end, gpu_allocate, gpu_free, gpu_get, gpu_put},
   sensitive = false,
   morecomment = [l][\color{blue}]{\#},
   basicstyle=\footnotesize\ttfamily
   }
\usepackage{url}
\usepackage{tabularx}

\begin{document}
	
\title{Leveraging the Super Instruction Architecture to Develop Massively Parallel Computational and Environmental Chemistry Applications
\thanks{
JNB, SEM, DSB, VFL were 
funded by an internal R\&D grant from ENSCO, Inc.  This research used resources of the Oak Ridge Leadership Computing Facility, which is a DOE Office of Science User Facility supported under Contract DE-AC05-00OR22725.  This material is based, in part, on research sponsored by AFRL/RXAP under grant FA8650-17-1-5401.     
The US Government is authorized to reproduce and distribute reprints for Governmental purposes notwithstanding the copyright notation thereon.  The views and conclusions contained herein are those of the authors and should not be interpreted as necessarily representing the official policies or endorsements, either expressed or implied, of the AFRL/RXAP or the US Government.}
}

\author{
\IEEEauthorblockN{Jason N. Byrd\IEEEauthorrefmark{1},
Stephen E. Masters\IEEEauthorrefmark{1},
Douglas S. Burns\IEEEauthorrefmark{1},
Victor F. Lotrich\IEEEauthorrefmark{1},  and 
Beverly A Sanders\IEEEauthorrefmark{2}
\IEEEauthorblockA{\IEEEauthorrefmark{1}ENSCO, Inc.,  Melbourne, FL, USA\\
Email:  byrd.jason@ensco.com, masters.steve@ensco.com,  burns.doug@ensco.com, lotrich.victor@ensco.com }
\IEEEauthorblockA{\IEEEauthorrefmark{2}Department of Computer \& Information Science \& Engineering,  University of Florida,  Gainesville, FL, USA\\
Email: sanders@cise.ufl.edu
}
}
}

\maketitle

\begin{abstract}
The task of developing high performing parallel software must be made easier and more cost effective in order to fully exploit existing and emerging large scale computer systems for the advancement of science. The Super Instruction Architecture is a parallel programming platform geared towards applications that need to manage large amounts of data stored in potentially sparse multidimensional arrays during calculations. The SIA platform was originally designed for the Quantum Chemistry software package ACESIII. More recently, the SIA was reimplemented to overcome limitations in the original ACESIII program. It has now been successfully employed in the new Aces4 Quantum Chemistry software package and the development of the atmospheric transport application MATLOC, thus demonstrating the versatility of the SIA approach. MATLOC calculates transport and dispersion of mass over regions in the range of 100-1000s of square kilometers and is a significant improvement over existing community codes. This paper describes results from both the transport and dispersion application as well as some difficult Quantum Chemistry open shell coupled cluster benchmark calculations using Aces4.
\end{abstract}

\section{Introduction}

	

To fully utilize emerging compute architectures it is critical that the development of high performance parallel software be made as rapid and cost effective as possible. To this end it is  important to develop maintainable and robust computational programs that can be easily ported to a wide range of computing architectures.  One strategy to achieving maintainable, robust and yet efficient and scalable software is to develop the specific domain models using a shared code base and framework.  Such a framework should be responsible for interfacing with the machine, handling data I/O and communication, and yet be flexible enough to be used by technical experts from unrelated domains.
 The development of the Super Instruction Architecture (SIA) \cite{lotrich2005,lotrich2008chem,Sanders:2010:BLR:1884643.1884677} aims to cultivate a high-performance computing code development process and software stack that addresses the needs of the general scientific modeling community, especially targeting scientific applications with extremely large arrays. The SIA is realized in a parallel programming platform that comprises a domain specific language (DSL), SIAL, and its runtime system.  
In this paper we describe the use of the SIA to develop highly scalable and efficient implementations within two completely different scientific domains, each with its own particular computational bottlenecks and requirements. 
Quantum Chemistry programs are predominantly computationally intensive applications with difficult I/O requirements when out-of-core algorithms are used. Conversely, transport and dispersion (T\&D) applications are entirely I/O and memory dominant algorithms that require very efficient data management to be effective. This work demonstrates that the SIA is flexible enough to act as a computational kernel DSL platform, and as a complex data management layer depending on the demands of the domain.

Computational chemistry is a mainstay tool in any high-fidelity material science.  
The SIA was originally conceived as a parallel programming platform for quantum chemistry, in particular highly accurate many-body solutions 
to the Schr\"{o}dinger equation. Its original instantiation was ACESIII (Advanced Concepts in Electronic Structure)  \cite{acesiii2008}, a
software package for 
highly accurate \textit{ab initio} electronic 
structure methods. Its successor, Aces4 \cite{aces4github, aces4IPDPS17, exascale2018}, was
 a complete from-the-ground-up reimplementation including   
extensions to the SIAL language, important
new capabilities to the underling architecture such as support for block-sparse arrays, an improved software architecture, and implementations of new methodological and implementation developments enabled by the enhancements.

Every year, the demands set forth by the community, be it academic, industrial, or US Government (e.g. DOE, DOD, DHS)
have pushed the size of the physical problem to be studied to larger and larger systems such as surfaces, nanoparticles and crystals.  Standard 
computational tools used in performing many calculations have fixed algorithmic scaling (high fidelity methods are polynomial in scaling) and 
there is no alternative unless fidelity and reliability are sacrificed.  These computations are typically unfeasible with standard methods and 
difficult or impossible to implement with most existing software platforms because they can generate hundreds of terrabytes of data and have 
scaling characteristics that prove challenging for developing efficient implementations.  Aces4 was designed with such scalability and performance demands in mind.
A well-established approach to demonstrating code performance in computational chemistry is to perform a benchmark calculation 
of a desired molecular property (typically the total energy) for very large molecules.  Historically, the methods used to perform benchmark calculations have  been coupled-cluster singles and doubles with perturbative triples (CCSD(T), an $O(n^7)$ algorithm) restricted to closed shell molecules.  In this paper, we present an open shell (some unpaired electrons) benchmark, which is at least three times more computationally  expensive, on a size of molecular systems never studied with this level of theory.

Although the motivating domain was Quantum Chemistry, an obvious question arises as to whether the approach is more generally useful.  
In this paper, we demonstrate that the answer is a resounding yes.  
Modeling the impact of atmospheric chemical and particulate material, such as industrial pollution or a material from a natural disaster, is 
an important tool for first responders, monitoring agencies, and policy makers to name a few.	The SIA was recently used to 
implement an application in this domain.  We describe MATLOC (MAssively parallel Thread LOcation Capability), which
calculates mass dispersion and 
transport over regions in the range
of 100-1000s of square kilometers with a computational performance that is a significant improvement over existing community codes. 
Scaling benchmarks using the Fukushima test set will be presented.

The contributions of this paper are
\begin{itemize}
\item A description of the novel climatological T\&D MATLOC package, the first significant application of the SIA in a new domain. MATLOC makes mass transport studies that were previously computationally impractical routine.
\item Demonstration of MATLOC's linear scaling implementation and overall performance when applied to a study of unprecedented size.
\item Head to head scaling comparison between the open shell CCSD(T) massively parallel community standard package ACESIII and the new Aces4/SIA implementation showing a $20-30\%$ performance improvement of Aces4 over ACESIII across most processor and job size ranges.
\item Showcase of Aces4/SIA's open shell implementation by executing some of the largest open shell CCSD(T) calculations ever using the US Department of Energy's leadership class platform Titan.
\end{itemize}

\section{Overview of the Super Instruction Architecture}
In this section, we give a very brief overview of the SIA. It has been described in more detail, along with the quantum chemistry application Aces4
in\cite{aces4IPDPS17,exascale2018}.

The key abstraction of the SIA is the notion that multi-dimensional arrays are partitioned into blocks (also called tiles).  This is, of course, not a new idea, but the way that the blocks are referenced in the DSL, SIAL,  and managed by the runtime system, is unique in the SIA.  To define an array, one first defines a set of index types, each with a maximum range and a partitioning of the range into segments.  A multidimensional array is then defined by specifying the index type for each dimension and  large multi-dimensional arrays are partitioned into blocks as implied by the segmentation of their indices.  The segments within an index are not necessarily of uniform size, thus neither are the blocks.  This is important because
in quantum chemistry, the segmentation choices are driven by both performance considerations and the physics of the problem under consideration.
Blocks are referenced within SIAL using segment numbers and algorithms are expressed in terms of operations on blocks rather than on individual floating point numbers.  A rich and \textit{extensible} set of operations on blocks is provided. 
These operations, or computational kernels, 
are called super instructions which operate on 
super numbers (blocks), hence the name Super Instruction Architecture.
Some super instructions are built in to the DSL; others are provided by domain scientists. 

The SIA runtime is implemented in C++ using MPI.\footnote{A non-parallel version without MPI is also maintained. It is sometimes convenient for application developers,
but also serves to demonstrate that the dependence on MPI is sufficiently modularized that it could be easily replaced with a different communication
layer without requiring any changes to the application code, and few change to the SIA itself.}
Each MPI process is either a server or a worker.  
The main task of a server process is to store and serve blocks of distributed arrays that have been assigned to it.  This also includes responding
to worker requests to send or update blocks.   Servers can also perform 
simple computations on single blocks such as initializing all the elements of a block to a given value or atomically accumulate blocks arriving from workers
into existing blocks.    
The SIA is specifically intended for calculations that involve very large amounts of data; it is quite possible for data generated during a calculation 
to exceed available memory.  Thus servers can write blocks to disk in order to free memory and restore
them when needed using special purpose memory and disk file layouts specialized for the SIA along
with protocols implemented using MPI-IO. Except for the increased latency, this functionality is transparent to the workers (and the SIAL programmer). 

SIAL programs are compiled into SIAL bytecode which is interpreted by worker processes to perform the actual calculation.
At any point in time, the worker may hold blocks from distributed arrays which are transmitted to and from servers as required.    
Each such block is either a copy of a block held at its owning server, or
an updated version that will replace or be accumulated into a block at its owning server.  The blocks are held in a specialized map along with blocks from
small replicated and local arrays.  A block's metadata includes
its shape and status; for example, whether the block is part of a pending asynchronous operation.



\begin{figure}
\begin{lstlisting}[basicstyle=\footnotesize, numbers=left, numberstyle=\tiny, numbersep=5pt, escapeinside={/*@}{@*/},columns=fullflexible]
pardo M,N,I,J  #parallel loop over ranges of indices M,N,I,J /*@\label{code:pardo1}@*/
  tmpsum[M,N,I,J] = 0.0  #assign 0.0 to all element of the block /*@\label{code:initialize}@*/
  do L  #serial loop over the segments in index L /*@\label{code:serial-loop-L}@*/
    do S  # serial loop of the segements in index S /*@\label{code:serial-loop-S}@*/
      get T[L,S,I,J] #request block of array T from its owning server /*@\label{code:firstLoopLine}@*/
      get V[M,N,L,S] #request  block of array V from its owning server /*@\label{code:integral}@*/
      tmp[M,N,I,J] =   V[M,N,L,S] * T[L,S,I,J]  #block constraction /*@\label{code:contract1}@*/
        #the system will wait, if necessesary, for requested blocks to arrive
      tmpsum[M,N,I,J] += tmp[M,N,I,J]  # elementwise accumulation /*@\label{code:tmpsum}@*/
    enddo S
  enddo L
  put R[M,N,I,J] = tmpsum[M,N,I,J] #replace  block of R on owning server with the contents of local block tmpsum /*@\label{code:put}@*/
endpardo M,N,I,J
sip_barrier #wait for all workers to arrive here /*@\label{code:sip_barrier}@*/
\end{lstlisting}
\caption{SIAL implementation of tensor contraction $
R^{\mu\nu}_{ij}
  =  \sum_{\lambda\sigma}
        V^{\mu\nu}_{\lambda\sigma} T^{\lambda\sigma}_{ij} 
$.  Index and array declarations have been omitted.
}
\label{fig:CCterm}
\end{figure}

The SIAL code fragment
in Fig. \ref{fig:CCterm} computes a four-index tensor contraction and gives an idea of SIAL functionality.
\lstinline|T|, \lstinline|V|, and \lstinline|R| are distributed arrays.\footnote{The necessary declarations of arrays and indices are not shown.}
The range of each dimension is partitioned into segments; 
\lstinline|M,N,I,J,L|, and \lstinline|S|
are indices that count segments rather than individual elements.  
Thus  \mbox{\lstinline|T[L,S,I,J]|}, \mbox{\lstinline|V[M,N,L,S]|}, and \mbox{\lstinline|R[M,N,I,J]|} 
are blocks of the distributed arrays, not individual floating point numbers.  
\mbox{\lstinline|tmpsum[M,N,I,J]|}, and \mbox{\lstinline |tmp[M,N,I,J]|} are appropriately sized temporary blocks 
automatically allocated when referenced and deallocated when no longer in scope.  
%
%

A SIAL program expresses the 
coarse-grained structure of the parallel computation. 
The details necessary for robust, high performance calculations are managed by the runtime system.  
Among these include memory management, performing asynchronous communication allowing computation and communication to overlap,  load balancing, and check pointing. A typical calculation  is comprised of several steps each implemented with a separate SIAL program.  A convenient mechanism for passing arrays between SIAL programs is available. 

The domain scientist will typically also need to provide 
implementations for some user-defined super instructions.  
These super instructions  take a set of blocks as input 
and update and/or generate new blocks as output.  They do not engage in communication.
Super instructions may be implemented 
in almost any convenient general purpose programming language;  
the only requirement is that procedures in the language interoperate with C++. 
Super instructions may need to be able to invoke C++ in order to request certain services from the runtime system, such as memory allocation and deallocation.

Communication with
servers uses asynchronous computation.  For example, the implementation of \mbox{\lstinline|get T[L,S,I,J]|}
sends a message requesting the block to the appropriate server, then posts an asynchronous receive.  Computation continues, and
when the block is needed in line \ref{code:contract1}, receive status is checked and the worker waits if necessary.  
One can often arrange things so that more work is done between get and the use of the block,
thus overlapping communication with computation.   
Similarly, the implementation of   \mbox{\lstinline|put R[M,N,I,J] = tmpsum[M,N,I,J]|}  in line \ref{code:put} performs an asynchronous protocol to transfer the block to the server.  The runtime system
ensures that the block \mbox{\lstinline|tmpsum[M,N,I,J]|} is not overwritten or deleted until it is safe to do so. In the meantime, the worker continues with subsequent commands. 

Some important aspects of the way blocks are represented and accessed include:
\begin{enumerate}
\item Each block has a convenient identifier which serves as a key into specialized maps.   
\item No memory for data or metadata of a block is allocated (except for blocks of so-called static arrays, which are replicated at all processes and should be small) until it is created and used in the calculation.  
\item At each process, block-specific metadata only exists for blocks that are currently located in that 
process.
\item Block IDs are maintained and known to the runtime system wherever a version of a block is located. 
\item
The SIAL compiler deduces how blocks will be accessed (read, write, update, accumulate into, etc.) and encodes this information in the byte-code.  
\end{enumerate}
The second and third items above are key to efficiently exploiting sparse arrays.
The latter two items, along with a small amount of metadata per block, make it possible for the runtime system to perform several valuable functions including
ensuring consistency of multiple copies of a block, 
performing inexpensive
runtime checks for data races at servers, and
transparently caching copies of blocks until either they are no longer valid or the memory is required.  
All of these actions occur at the granularity of a block, amortizing the overhead over a large number of elements. 

The \lstinline|sip_barrier| instruction on line \ref{code:sip_barrier} of Figure \ref{fig:CCterm} is the only synchronization primitive in SIAL.  Semantically, it means that all operations before the barrier complete before any of the the operations after the barrier are executed. 
In contrast, for example to OpenMP,
there are no implicit barriers at the end of loops, rather they are all explicitly given in the SIAL program.  This means that if a SIAL program contains two parallel loops with no 
barrier in between, a worker will immediately proceed to the subsequent loop after finishing the
first loop.  
Barriers naturally divide the computation into a sequence of numbered phases agreed upon by all workers, with the property that the sending phase of messages received at all servers is monotone.  A data race occurs if two accesses in the same phase conflict:  i.e. they originate from different workers and at least one modifies the block in a way that is not an accumulate.  
Information about the last access to a block at a server, including access type, worker, and phase number is recorded, allowing data races to be dynamically detected with negligible overhead.
In contrast, one race detection tool for general MPI one-sided communication\cite{Chen:2014:MDM:2683593.2683648} reported an average of 45\% overhead.  Experience has shown that data race detection in the SIA provides valuable
information during application program development.

To execute an application program built using the SIA, input data in a convenient format
is given to a domain specific pre-processor.  The Aces4 preprocessor, for example, parses a so-called ZMAT file containing
the molecular geometries, a basis set specification, and the job flow specification.  It generates
a file containing the list of SIAL programs to execute, information about array sizes and segmentation, and initial values 
for predefined constants and arrays in the required format. 
MATLOC has a similar preprocessor 
which extracts the required data tensors from output from the 
standard community 
program Weather Research and Forcasting (WRF) v3.9\cite{WRF39}\footnote{Only $\sim33\%$ of the total tensors in 
a WRF output file are required by MATLOC.}. Setting up the grid of potential sources, defining or computing the various conformal grid variables, 
and assigning releases to specific segments are tasks also handled by the pre-processor.

The SIA has been used on a variety of machines including  HiPerGator @ University
of Florida, DOD Garnet Cray XE6, DOD Haise IBM iDataPlex, DOD Armstrong Cray
XC30, DOD Excalibur Cray XC40, DOD Topaz SGI Ice X, DOE Titan Cray XK7.  For
convenience of developers, it also can be run on OSX 10.9 and greater.  

\section{Related work} 
Many applications that deal with large, partitioned, arrays 
use some sort of middleware.  
Perhaps the most well-known middleware for dealing with large distributed arrays 
is the Global Arrays (GA) toolkit \cite{harrison1993}. GA was used to 
implement NWChem\cite{nwchem2010}, one of the most prominent  
parallel quantum chemistry suites, and has been used in many other application domains.
Global arrays statically partitions the data belonging to a dense array among the
processes participating in a calculation and 
offers a convenient programming model with a one-sided abstraction
that allows arbitrary sections of arrays to be
accessed in a shared memory-like style using a global index space.  
The basic programming model is to copy a section of an array from GA to local memory (get), 
update it locally, then copy or accumulate it back in global memory (put).  
Programmers need to manage consistency explicitly using a sync operation.
The GA toolkit has been in development for more than 20 years and offers 
a variety of enhancements to the basic functionality described above. 

The SIA programming model is based on blocked arrays and blocks are the object of interest rather than individual array elements, or ranges of individual array elements.
Generally there are many more blocks than servers. 
Also, memory is allocated for a block only when the block is used, providing a way to handle block-sparse arrays.  
Although a recent enhancement of
GAs added the option for block-cyclic distribution (whose use disables some of GAs features), traditionally arrays were partitioned among 
processes with each process having (at most) one section of an array. 
In the SIA, the interface for obtaining a block of an array requires indicating the access mode, which along with 
maintenance of identity of blocks allows the SIAL compiler and runtime system
to provide significant error checking.  It also enables functionality such as interoperability between local, contiguously stored and
distributed arrays along with maintenance of consistency of blocks. In GA, after a segment of
an array is copied to a local buffer, the GA system loses track of its relationship with its source.  
In the SIA, distributed arrays are managed by server processes, while computation is done by worker processes.  This is in contrast with GA, where,
as typically used, all processes are workers and each owns part of the array.  
SIA distributed arrays are supported by a special purpose virtual memory system, which is integrated with checkpointing, and 
that writes blocks to disk
when memory is required at a server, reads them back in when required, transparently.
A library distributed with GA, Disk Resident Arrays (DRA) allows arbitrary segments of global arrays to be explicitly copied to and from disk. This is done using
collective IO operations, thus requires coordination between all workers.

A more recent development with similar goals is the TiledArray\cite{tiledarray} framework, also originally motivated by the needs of Quantum Chemistry.
TiledArray makes sophisticated use of C++ templates to offer general support for large multidimensional arrays,
including block-sparse arrays.  TiledArray arrays are statically partitioned 
into blocks called tiles which are distributed among all of the processes in the
computation.  Block-sparse arrays are handled by maintaining a bitset for
each array with a bit for each tile that indicates whether or not all elements 
in the tile are zero.  If all elements in the tile are zero, the tile representation does not
allocate memory for the tile's elements.  
TiledArray uses the MADNESS system's\cite{DBLP:journals/corr/HarrisonBBCFFGH15} parallel runtime, MADworld, as their underlying runtime system.  MADNESS dynamically creates tasks that
can execute when their inputs are available, thus conceptually forming a directed acyclic graph (DAG), whose edges represent dependencies.  Tensor expressions, which are fundamental to QC calculations,
are supported with a very small DSL that is embedded in C++.
This is an alternative approach to a DSL than provided by SIAL.  SIAL is a standalone programming 
language that manages the flow control of the computation while the TiledArray is restricted in scope to the domain of tensor expressions and is
embedded into C++;  programmers write programs in C++ which create TiledArray objects as part of their data structures, and invoke snippets of the DSL in 
the form of string parameters passed to operators defined on TiledArray objects.
MPQC \cite{peng2019}, a recent quantum chemistry package discussed in section \ref{results:aces4}, was built using TiledArrays.

\section{Computational Chemistry}

\begin{table}
	\begin{tabular}{rccccc}
		\hline
		\hline
		Basis set & aug-pVDZ & pVTZ & aug-pVTZ & pVQZ & aug-pVQZ \\
		Functions & 360 & 484 & 748 & 912 & 1340 \\
		\hline
		\hline
	\end{tabular}
	\caption{The number of basis functions for DVOH for each basis used.\label{dvohbasis}}
\end{table}

Computational chemistry programs focus primarily on solving the clamped nucleus many-body Schr\"{o}dinger equation, the primary challenge being correlating the fermion particles (electrons).  The Schr\"{o}dinger equation is typically expressed as
\begin{equation}\label{schr}
\hat{H}|\psi\rangle = E |\psi\rangle,
\end{equation}
where $\hat{H}$ is the many-body electronic Hamiltonian, $E$ is the eigenvalue energy of the many-body system, and $|\psi\rangle$ is the eigenvector wavefunction that spans the Hilbert space of $\hat{H}$.  When solving the many-body problem as a mean field (the Hartree-Fock solution), (\ref{schr}) becomes an eigenvalue problem with the standard $O(n^5)$ scaling (where $n$ is the number of basis functions, e.g. the size of the Hilbert space).  More accurate (and more complicated) solutions to (\ref{schr}) involve perturbative expansions of the wavefunction that include coupled excitations from the mean field, see Refs \cite{bartlett2007,shavitt2009} for a detailed overview.  As these methods are well documented in the literature we will forgo a detailed description of the mathematics. 

The gold standard method in quantum chemistry is coupled-cluster singles and doubles\cite{purvis1982}, denoted CCSD, with perturbative triples, (T)\cite{Ragavachari}. CCSD is a non-linear post-mean field solution with $O(n^6)$ scaling which is solved iteratively, typically requiring 10-15 iterations to reach convergence. The \emph{post hoc} (T) calculation is a post CCSD perturbative correction with $O(n^7)$ scaling. The fundamental mathematical kernels involved in computing a CCSD or (T) calculation are tensor (rank 2, 4 or 6) transposes and contractions. For reference, a single index transpose would be represented as
$A^{ab}_{ij} = (B^{ab}_{ji})^T$ while a double contraction is demonstrated in Table \ref{fig:CCterm}.
Here $A$ and $B$ are tensors and $a,b\cdots,i,j,\cdots$ are indices. 
An important aspect of our recent work to improve Aces4 was to optimize the specific combinations of all contractions and transpose operations necessary to perform CCSD and (T) calculations.

Historically benchmark CCSD(T) calculations have been restricted to closed shell molecules (all electrons are paired).  Many chemical systems of interest are indeed closed shell so benchmarks against these are pertinent.  However, very little attention has been given to writing and demonstrating efficient open shell implementations of CCSD(T).  One reason for overlooking open shell implementations is justified by the plethora of closed shell chemical systems still to be studied, but the biggest factor is often simply that open shell is harder and more expensive.  When computing a \emph{closed shell} CCSD iteration, there are two large four-dimensional arrays stored and one five-dimensional loop over the largest indexes.  In contrast, an \emph{open shell} (denoted with a U-) CCSD iteration has six large four-dimensional arrays stored and three five-dimensional loops over the largest indexes.\footnote{The exact number of total intermediates and loops is implementation specific, however most implementations will have these particularly large terms}  Similarly, open shell U-(T) has twice as much data and calculations of a corresponding closed shell implementation.  Since many chemical systems we are interested in involve the oxidation of atmospheric pollutants, efficiently computing U-CCSD gradients\footnote{A gradient refers to taking the derivative of the wavefunction as a function of atomic nucleus position.} and U-CCSD(T) energies is critically important. Note that U-CCSD gradients are not yet implemented in Aces4, but a surrogate to a single gradient calculation can be estimated by multiplying the U-CCSD time by three.  

\begin{figure}[b]
	\includegraphics[width=0.4\paperwidth]{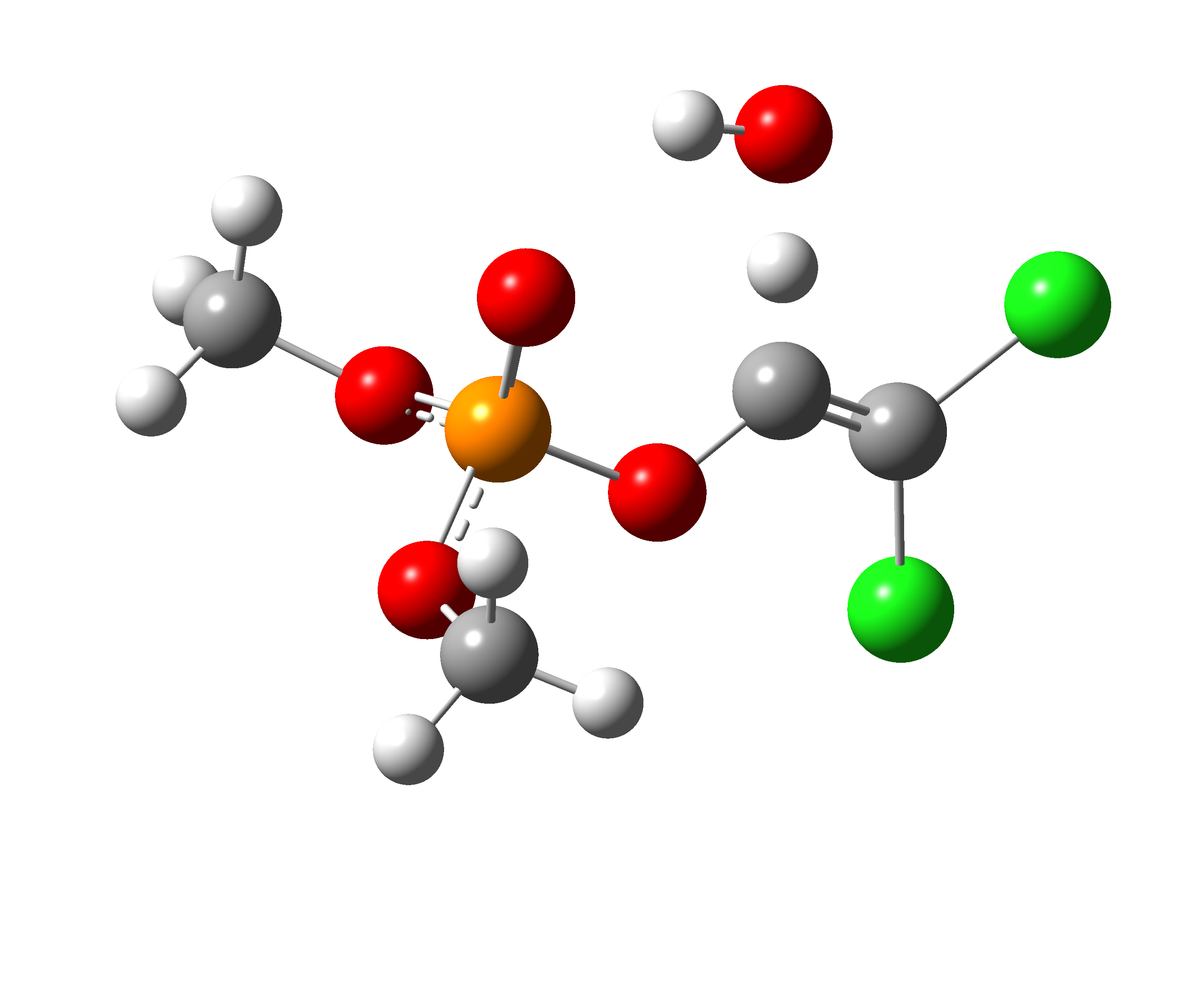}
	\caption{Schematic of dichlorvos reacting with a hydroxyl radical. 
		\label{dvohimage}
	}
\end{figure}
  
Dichlorvos, shown in  Fig. \ref{dvohimage}, is an insecticide who's application to farms and related areas via aerial distribution and environmental fate is of interest.  To investigate the computational scaling of Aces4's U-CCSD(T) implementation, we have performed a set of calculations on the open-shell compound dichlorvos + $\bullet$OH, denoted DVOH.  This is representative in size and structural similarity to  the calculations we perform routinely and  has  121 electrons and 20 atoms.   This is an open shell reaction with a key atmospheric oxidizer $\bullet$OH.

We have also performed some very large U-CCSD calculations to push the implementation and provide an estimation of how a gradient calculation would perform.  The first example is the biologically relevant two DNA base pair fragment GC-dDMP-B, guaninecytosine
deoxydinucleotide monophosphate in B-conformation with a sodium cation to provide a zero total charge, has been used as a case study for the closed shell performance of NWChem's CCSD(T) implementation \cite{nwchem2010,anisimov2014} and MPQC's \cite{MPQC} CCSD(T) density fitted approximation. (See \cite{peng2019} and citations therein for details of the approximation.)  GC-dDMP-B has 300 electrons and 63 atoms, which is about twice the size dichlorvos. The second very large U-CCSD calculation was performed on the explosive octogen, otherwise known as HMX. HMX has 152 electrons and 28 atoms. Understanding the environmental fate of HMX is quite important as unexploded ordinance poses an environmental hazard in addition to the obvious health problems. While GC-dDMP-B and HMX are technically closed shell compounds, modeling the dissociation of HMX involves stretched bonds where open shell models are required to get the right spin combinations. GC-dDMP-B is one of the few very large compounds with published CCSD timing information for comparison.  Details of the dichlorvos, GC-dDMP-B, and HMX calculations are described in section \ref{results}.

\section{Transport and Dispersion Modeling}

\begin{figure}
	\includegraphics[width=0.4\paperwidth]{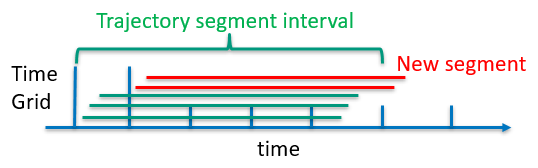}
	\caption{Segmentation approach for grouping mass releases and virtual sources. 
		\label{tdsegment}
	}
\end{figure}
Transport and dispersion (T \& D) models help determine (a) where a release of gas or particulates in the atmosphere will travel and (b) characterize how the concentration of the material will change as a function of time and distance.  This could be used, for example, to help first responders deal with chemical spills or radioactive accidents, by indicating which populations should be evacuated, and from which direction emergency personnel should approach to avoid exposure.  The models have been used to study and forcast a variety of materials such as radioactive material, wildfire smoke, windblown dust, pollutants, pesticides, and volcanic ash.  

MATLOC is a parallel T \& D program who's development was funded 
by ENSCO, Inc.
to address gaps in the available T\&D tools. 
Based loosely on a Lagrangian puff model\footnote{In this approach, the tracked material is modeled as a set of  ``puffs'' that expand via dispersion, causing the concentration of the material of interest to decrease.  Once the size of the puff becomes too large, it splits into several new puffs, each with its share of the mass of the material.}
developed for the US Air Force in 1989 and last updated in 2008 \cite{SLAM2008}, MATLOC was designed from the ground up to enable climatological studies of mass transport from local areas (tens to hundreds  $km^2$) to the regional scale (hundreds to thousands $km^2$) 
involving tens of thousands to millions of individual mass releases to provide a statistical representation of the climatological and spatial variations. There are many T\&D models in the community \cite{HOLMES20065902}, including several long standing benchmark codes such as SCIPUFF \cite{scipuff} and HYSPLIT \cite{stein2015,rolph2017}.  These codes excel at modeling a few hundred releases, using a mature code base that is typically OpenMP threaded to a few tens of processors.  Recently however, interest has been expressed by end users in the ability to perform studies on larger source areas, for longer periods of time, with atmospheric reactive chemistry (e.g. model not just the initial material but degredation compounds as well), and statistically combine the results over many years to create a climatological assessment, all while retaining the distinct information that links a particular mass measurement to a specific source location and release time (something most codes do not allow).  Estimates for performing an ensemble study of a farm land region's pesticide use, where one would model $100$ farms for a whole year for a $10$ year climatology could result in $\sim50$ million distinct releases requiring $20-40$ Tb of input data and $100s$ of Tb of output data with a wallclock time running into the tens of months.
The limiting factor is typically not the physics or computational algorithm but in the way the models were originally designed to manage the increasing amount of data.  

Lagrangian puff models consider a distinct unit of mass in the atmosphere that has a spatial concentration profile, $C$, represented by the product of horizontal and vertical Gaussians \cite{pasquill1971},
\begin{multline}
C(t,x_0, y_0, z_0,\sigma_h,\sigma_z,m)=\\
\frac{m}{(2\pi)^{3/2}\sigma_H(\bar{x},t)^2\sigma_z(\bar{x},t)^2}\exp\lbrace-\frac{(z-z_0(t))^2}{2\sigma_z(\bar{x},t)^2}\rbrace\times \\
\exp\lbrace-\frac{(x-x_0(t))^2+(y-y_0(t))^2)}{2\sigma_H(\bar{x},t)^2}\rbrace,
\end{multline}
where $m$ is the total mass in the puff, $x_0(t)$, $y_0(t)$, $z_0(t)$ are the puff's center coordinates,  $\sigma_x = \sigma_y = \sigma_H(\bar{x},t)$ and $\sigma_z(\bar{x},t)$ are the puff's temporally and spatially varying horizontal and vertical distribution functions respectively.  The puff distribution functions, $\sigma_H$ and $\sigma_z$, approximate molecular diffusion and atmospheric turbulence using empirically derived functions \cite{SLAM2008}.  Puffs exist within vertical layers described by $z_{\rm min}$ and $z_{\rm max}$, where the layer boundaries are defined by specific situations dictated by the surface of the ground and the height of the planetary boundary layer\footnote{The planetary boundary layer is a more turbulent atmospheric layer where the atmosphere is more strongly influenced by surface roughness and solar radiation effects.}  $Z_{\rm PBL}$.   
As puffs propagate, they will grow wider ($\sigma_H$) and taller up to $Z_{\rm PBL}$.  Should a puff grow too wide, the model vertically splits the puff into $n$ distinct puffs in an approximation to the original Gaussian distribution (while exactly conserving mass). This is particularly important when modeling in mountainous terrain or when the wind is highly turbulent.  Should the $Z_{\rm PBL}$ height decrease or increase (due to thermal effects in the atmosphere increasing or decreasing turbulence) then the puff will continue to grow up to $Z_{\rm PBL}$, or be horizontally split into two puffs whose new $z_{\rm max}$ and $z'_{\rm min}$ will be $Z_{\rm PBL}$.  Horizontal splitting accounts for sheering of the mass due to wind field speed and direction difference above and below the planetary boundary layer.  The consequence of horizontal and vertical splitting is that as a puff is propagated through time, the initial mass may end up splitting into a large number of smaller puffs each of which must be tracked.  

\begin{figure}
	\includegraphics[width=0.4\paperwidth]{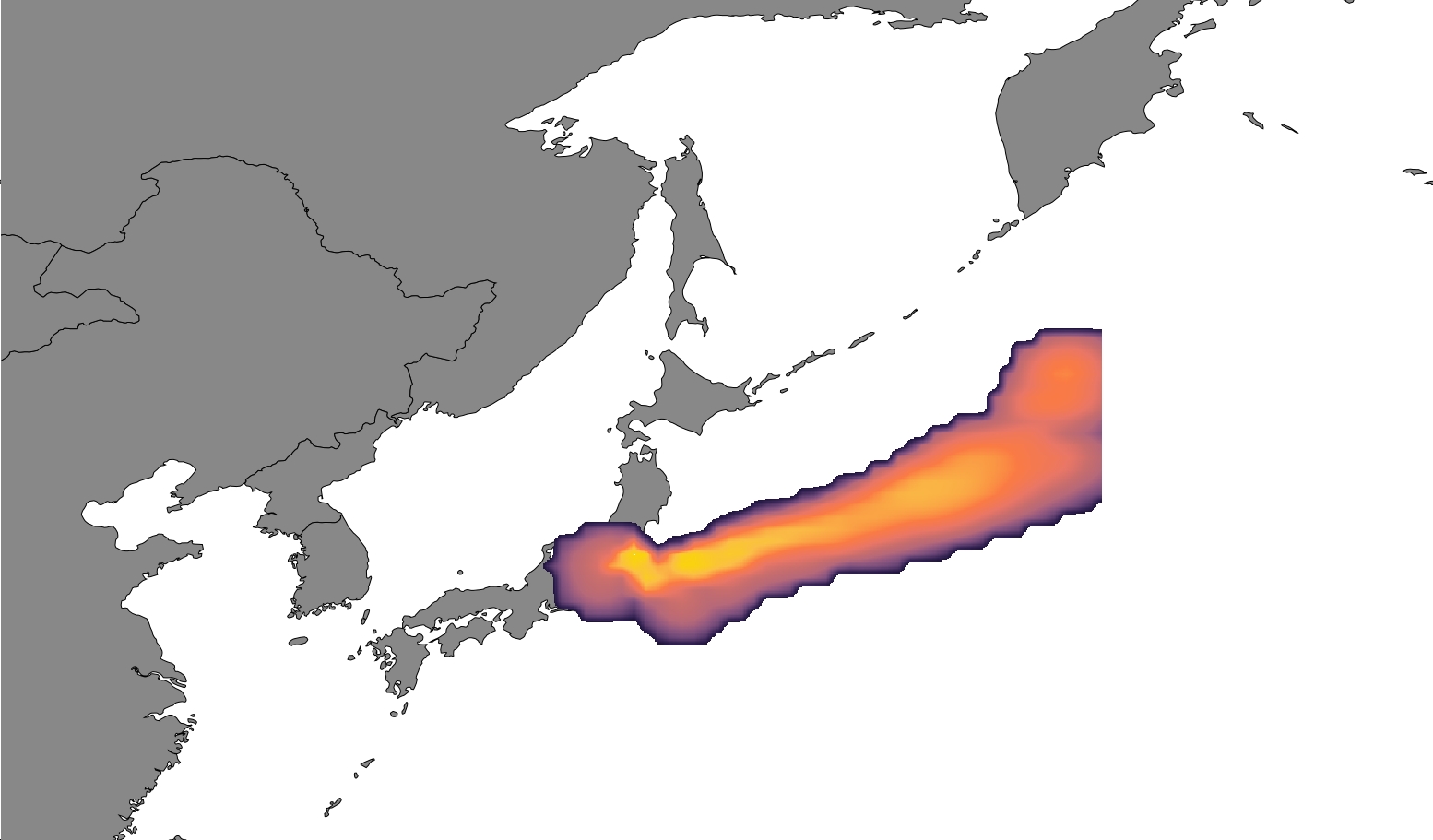}
	\caption{Illustrated puff (6 hour integrated) originating from Fukushima Japan. 
		\label{fukushima}
	}
\end{figure}

The center of the puff, $\bar{q}(t)=\lbrace x_0(t),y_0(t)\rbrace$ is propagated horizontally by
\begin{equation}\label{tdprop}
\frac{d}{dt}\bar{q}(t)\simeq \bar{F}(\bar{q}(t),z,t)
\end{equation}
where $\bar{F}(x,y,z,t)$ is the wind vector field supplied by a numerical weather prediction model (NWP).  The preprocessor interprets netCDF output from the standard community NWP program Weather Research and Forcasting (WRF) v3.9 \cite{WRF39}. A single propagation step of the puff center is performed using a two-step predictor-corrector approximation to (\ref{tdprop})
\begin{align}\label{pcprop1}
\bar{q}'(t+\Delta t) &= \bar{q}(t) + \bar{F}(\bar{q}(t),t) \\
\label{pcprop2}
\bar{q}(t+\Delta t) &\simeq \bar{q}(t) + \frac{1}{2}\left( \bar{F}(\bar{q}(t),t) + \bar{F}(\bar{q}'(t+\Delta t),t+\Delta t)\right) \Delta t.
\end{align}
The wind field $\bar{F}$ is provided by WRF as a gridded tensor which is discretized in the $x,y,z$ and $t$ dimensions.  In order to perform a single step with (\ref{tdprop}) it is first necessary to take the discretized wind field $\bar{F}(x_i,y_j,z_k,t_\ell)$ and obtain the tensor elements $\{x_i,y_j,t_\ell\}$ that bracket the desired point $x,y,t$ and vertically average the field
\begin{equation}
\bar{F}(x_i,y_j,t_\ell) = \frac
{\int_{z_{\rm min}}^{z_{\rm max}} \bar{F}(x_i,y_j,z,t_\ell) dz}
{z_{\rm max}-z_{\rm min}}.
\end{equation}
We use a linear interpolation for time followed by a bi-linear interpolation in space to obtain each $\bar{F}(\bar{q}(t),t)$ used in Equations \ref{pcprop1}-\ref{pcprop2}.

Puffs are propagated through time to some predefined max (often 24 hours) and stored as trajectories which retain the historical position and other parameters of the puff,  as well as any splits. 
MATLOC needs to use five three-dimensional (namely the ${\bf x}$ and ${\bf y}$ vector fields, WRF sublayer height above ground, pressure, and temperature arrays), and nine two-dimensional tensors (including the surface pressure, boundary layer height above ground, etc.) from WRF to perform a calculation and provide all the requisite data for down-stream codes.  A unique requirement demanded of MATLOC is to retain the individual history of each puff such that each puff release is treated entirely separately from any other release. This is significantly different to other models such as SCIPUFF.  While retaining each trajectory independently from the others increases the overall data space (in SCIPUFF puffs can merge, reducing total footprint), it does force the computational ($O$) and I/O ($K$) scaling to 
\begin{equation}\label{beforescaling}
O(n) + K(n)
\end{equation}
where $n$ is the total number of releases in the simulation. A linearly scaling algorithm (and we demonstrate in Section \ref{results} that the code is also linearly scaling) is perfect for an embarrassingly parallel implementation. However, the overhead of $K(n)$ becomes exceedingly high when performing climatology studies with high-fidelity WRF data.  Instead we group each release for each virtual source into a segment (see Figure \ref{tdsegment}) which divides the $n$ releases into $N$ segments containing $\nu$ releases (such that $n=\nu N$), reducing  (\ref{beforescaling}) to
\begin{equation}\label{matlocscale}
O(n)+K(n/\nu).
\end{equation}
By appropriately choosing segment sizes, so that $\nu\sim 2000$, it is possibly to take $10^6$ releases and reduce the data I/O and message passing by $\sim 2000$. 
This segmentation approach lends itself perfectly to the way the SIA chunks and manages arrays; indeed the SIA entirely enabled the creation of MATLOC within the time and budget constraints imposed.  


To illustrate the performance of the parallelization of MATLOC, we have performed a series trajectory calculations originating in the Fukushima province of Japan (see Figure \ref{fukushima}). A WRF model run, centered on the province, was executed for an eight week period using four nested grids.  The outer most grid covers approximately nine million square kilometers, with output every 3 hours, while the inner most grid covered a few thousand square kilometers with 15 minute output.  In all, there are 120,000 tensors to read and store, totaling  $\sim 900$ GB of data.  This WRF configuration would be considered a high fidelity setup compared to standard calculations performed by our group, but certainly not outside the bounds of what could be expected to become routine in the future.  We have performed tracer-only releases, forgoing computing atmospheric chemistry on the released mass.  Further details are described in the Section \ref{results}.

\section{Results}\label{results}
\begin{figure}
	\includegraphics[width=0.4\paperwidth]{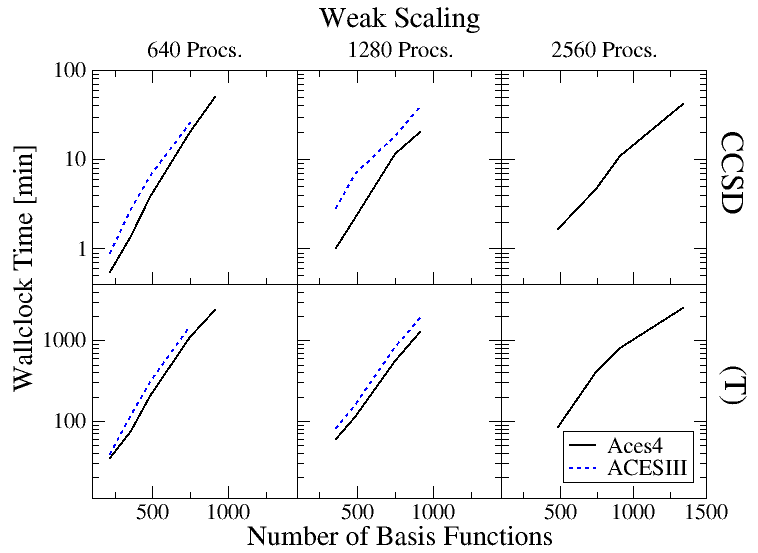}
	\caption{Weak scaling of a single DVOH CCSD(T) energy for a range of processors as a function of basis functions.  Loosely, the primary independent variable in the $O(n^6)$ and $O(n^7)$ formal scaling of CCSD and (T) respectively is the number of (AO) basis functions describing the linear algebra space.  The top row is the CCSD weak scaling per iteration, while the bottom row is the (T) weak scaling. 
		\label{weak}
	}
\end{figure}

\begin{figure}
	\includegraphics[width=0.4\paperwidth]{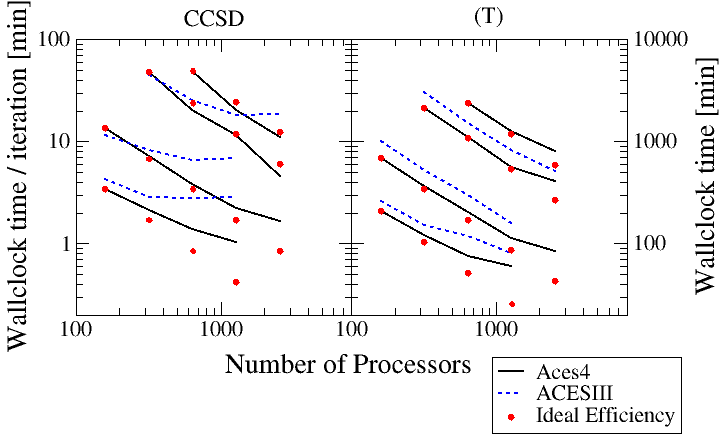}
	\caption{Strong scaling of wallclock times per iteration for a DVOH U-CCSD ($O(n^6)$) energy (left) and total wallclock time for a U-(T) ($O(n^7)$) energy (right).  The solid black line is from Aces4 while the dashed blue line is ACESIII, and the red dots represent the theoretically perfect efficiency scaling.  The basis functions used for each group of curves are, in order from bottom to top, aug-pVDZ, pVTZ, aug-pVTZ, and pVQZ (see Table \ref{dvohbasis} for more details).
Our group routinely performs calculations using the aug-pVDZ to pVTZ basis sets on 500-900 processors, with which Aces4 appears to provide 20-30\% improvement in efficiency allowing larger and/or more studies to be performed. 
		\label{strong}
	}
\end{figure}

\begin{table}
	\begin{tabular}{llllll}
		\hline
		\hline
		Compound & S/W & Machine & Nodes & Proc. & CCSD \\
		 &  &  &  &  & iter/min \\
		\hline
		HMX       & Aces4                      & Titan       & 600    & 2400    & 66  \\
		HMX       & Aces4                      & Titan       & 600    & 4800    & 56  \\
		HMX       & Aces4                      & Titan       & 1200   & 9600    & 31  \\
		GC-dDMP-B & Aces4                      & Titan       & 2400   & 19200  & 102  \\
		GC-dDMP-B & NWChem                     & Blue Waters & 1100   & 1100    & 72  \\
		GC-dDMP-B & NWChem                     & Blue Waters & 20000  & 20,000  & 13  \\
		GC-dDMP-B & MPQC                       & BlueRidge   & 64     & 1024    & 43  \\
		\hline
		\hline
	\end{tabular}
	\caption{Open shell CCSD wallclock time per iteration for the two very large examples, GC-dDMP-B and HMX. Listed for comparison are the recent closed shell CCSD NWChem and MPQC results for GC-dDMP-B. \label{largeccsdtable}}
\end{table}

All computational chemistry jobs were performed using the ACESIII and Aces4 programs on the US Army Research Laboratory Cray XC40 {\em Excalibur}  unless otherwise specified. Calculations on {\em Excalibur} were performed using half a node (16/32 processors) in order to increase the amount of memory per processor, so for example a 640 processor job involved requesting 1280 processors (40 nodes). T\&D calculations were performed on the US Army Research Laboratory SGI ICE XA {\em Centennial} using half the processors (20/40) on the "big memory" nodes which provide 512 GBytes per node (a standard node on {\em Centennial} has 128 GBytes). Calculations performed on {\em Titan}, the US Department of Energy leadership class Cray XK7, used half a node (8/16 processors) in order to increase the amount of memory per core. Every wallclock data point described in this section is an average of at least three separate calculations which begins to remove the variability of I/O and network load etc. on the system unless otherwise specified. For reference the typical standard deviation of wallclock duration is $1-2\%$ but can range up to $10\%$.  Details of machine configurations are provided in Table \ref{table:platforms}.

\subsection{Aces4 performance}  \label{results:aces4}


\begin{figure}[b]
	\includegraphics[width=0.4\paperwidth]{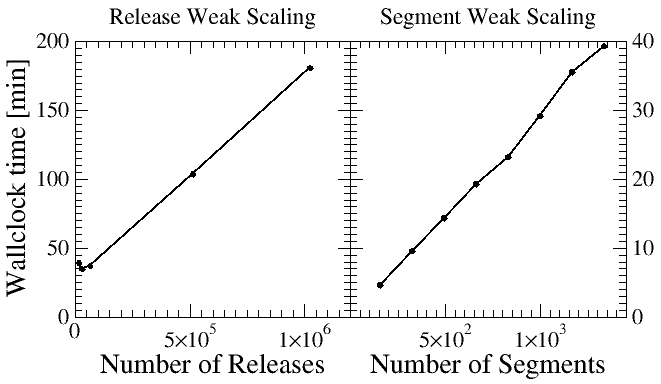}
	\caption{Weak scaling of MATLOC as a function of the total number of releases $n$, with the number of segments $N$ fixed (left), and number of segments $N$, holding the number of releases per segment $\nu$ fixed (right). 
		\label{td_combined}
	}
\end{figure}


The Aces4 U-CCSD(T) computational scaling data was computed using the DVOH molecule, chosen for its size and relevance to environmental chemistry. All quantum chemistry calculations on DVOH were performed using the standard Dunning \cite{dunning1989} basis functions with $24$ core orbitals dropped. As we are only interested in scaling data, we performed only four iterations of CCSD before using those amplitudes in (T). Table \ref{dvohbasis} relates number of basis functions to basis set used for DVOH. Weak scaling (constant number of processors, vary the number of basis functions) curves were generated for DVOH CCSD(T) energies using $640$, and $1280$ processors for Aces4 and ACESIII while only Aces4 was executed on 2560 processors due to allocation limitations. The processor range of $160$ to $2560$ is generally easy to request in a normal queue on most DoD Supercomputing Resource Center (DSRC) machines. The weak scaling log-y plot in Figure \ref{weak} demonstrates a well behaved polynomial scaling with respect to amount of work. Figure \ref{weak} also shows that Aces4 is out performing its predecessor ACESIII in every calculation, especially in the $300-600$ basis function range where our group's typical calculations reside. 
Deviations from the straight line illustrate where disc and network I/O begin to play a role, most significantly in the (T) calculation performed on $2560$ processors and $1340$ basis functions where the amount of work and data is becoming significant. 

Strong scaling (vary the number of processors, constant number of basis functions) curves were also generated for DVOH CCSD(T) energies using Aces4 and ACESIII. The strong scaling data is plotted in Figure \ref{strong} using a log-log scale, where a straight line with a slope of $0.5$ represents ideal scaling (plotted as red dots in Figure \ref{strong}). The CCSD implementation in Aces4 is consistently significantly outperforming ACESIII in every case except in lower processor count jobs, which is to be expected as ACESIII was optimized for those processor counts during its initial development. It is important to note that we are using the default settings for both Aces4 and ACESIII in these scaling calculations, which illustrates why not only is ACESIII competitive at lower processor counts but why it does not scale using more processors. The reason that Aces4's re-implementation of CCSD in SIAL is performant across the whole range of processors is due in part to changes in the current SIA implementation including the elimination of block wait time using asynchronous communication and significant improvements in how the meta-data for each block is handled, allowing more blocks to be instantiated without incurring a large overhead penalty. In contrast, the Aces4 (T) SIAL implementation is essentially a direct port of the ACESIII code. However while Aces4 and ACESIII scale essentially identically, Aces4 is consistently 20-30\% faster for every (T) energy computation, which highlights the performance improvements and has a direct impact on costs and performance of projects.


To stress test Aces4's U-CCSD implementation and illustrate this code's capability for larger molecular sizes and processor counts, we have applied it to HMX and GC-dDMP-B. We computed CCSD energies for HMX using the aug'-cc-pVTZ basis set (the prime denotes no diffuse basis functions on hydrogen) with $1032$ total basis functions and 20 core orbitals dropped.  For GC-dDMP-B we use the same basis set as Anisimov {\it et al.} \cite{anisimov2014}, namely the $6\text{-}311\text{++}G^{**}$ Pople basis set \cite{krishnan1980,mcLean1980} with $1081$ total basis functions and $47$ core orbitals dropped. Our results for HMX and GC-dDMP-B are presented in Table \ref{largeccsdtable} including comparisons from NWChem \cite{anisimov2014} and MPQC \cite{peng2019}. To provide a sense of scale, the most computationally intensive terms in CCSD scale $\sim o^2$, where $o$ is the number of occupied orbitals.  In this way GC-dDMP-B is approximately $4\times$ as intensive as HMX which is then $3\times$ larger than DVOH. Due to queue and allocation limitations the timings reported for Titan are single points and not averages of several runs. CCSD calculations for HMX were performed on Titan for a modest range of processors and demonstrate a $55\%$ scaling efficiency which is quite reasonable. The $4800$ processor wallclock value was obtained from a three iteration run, while the $9600$ processor calculation was allowed to run to convergence ($15$ iterations). 

Three U-CCSD iterations were performed for GC-dDMP-B, averaging $102$ minutes per iteration using $2400$ nodes on Titan.  This is contrasted with the closed shell CCSD wallclock time of $72$ ($13$) minutes per iteration performed using $1100$ ($20000$) nodes respectively on Blue Waters. Formally speaking open shell CCSD is $3\times$ as computationally intensive as the corresponding closed shell algorithm with $6\times$ the data. Keeping the formal scaling in mind, the recent Aces4 calculation compares very favorably to the NWChem calculations on a processor used basis.  It is important to also note that the number of resources consumed is an important metric to consider when developing massively parallel software, especially with the current push to develop exascale ready codes. In that context Aces4 performed very well against NWChem as those calculations in Table \ref{largeccsdtable} were limited to a single processor per node, so $20000$ processors required $20000$ nodes while Aces4 only needed $2400$ nodes to utilize $19200$ processors.

\subsection{MATLOC Performance}

In illustrating the scaling and performance of the T\&D program MATLOC, it is our first goal to demonstrate that it is in fact a linear scaling implementation as predicted by (\ref{matlocscale}). To accomplish this we have performed a weak scaling test where the number of releases per segment, $\nu$, is varied while keeping the segment count constant, or conversely vary the number of segments keeping the number of starts per segment the same. The results of both weak scaling curves are plotted in Figure \ref{td_combined}. Indeed, both plots show a very clear linear trend (aside from very low numbers of releases which illustrates the general overhead of the problem). It should be noted that performing T\&D calculations with $10^{6}$ releases is not routine, and in fact we are not aware of any group that can even perform that many at all without utilizing thousands of processors (at least within reasonable time frames). 

With such a clear embarrassingly parallel algorithm and implementation, MATLOC would be expected to also scale well for more processors so long as the inter-node communication backbone of the computer is performant and the amount of output data remains bounded.  During a recent field trial workup, performed well after the data in Figure \ref{td_combined} was obtained, we in fact reached a point where the number of output files in MATLOC became untenable for standard Lustre file systems. To better tune the output we combined the output files to a single file per segment, which lends itself very naturally for a simplistic but effective parallel I/O where every worker writes to a single file. Details of these improvements and related work will be documented in future publication.

\begin{table*}[tp]%
\caption{The hardware and software configuration of each machine used to produce results reported in the paper}
\label{table:platforms}
\begin{tabularx}{\textwidth}{
										| >{\raggedright\arraybackslash\hsize=.5\hsize\linewidth\hsize}X
                                        | >{\raggedright\arraybackslash\hsize=1.4\hsize\linewidth\hsize}X
                                        | >{\raggedright\arraybackslash\hsize=.6\hsize\linewidth\hsize}X
                                        | >{\raggedright\arraybackslash\hsize=1.5\hsize\linewidth\hsize}X 
                                        | >{\raggedright\arraybackslash\hsize=1\hsize\linewidth\hsize}X 
                                        | }
\hline
\bfseries{Machine	}		&	\bfseries{Location}			& \bfseries{Vendor and configuration}		&	\bfseries{Linux Kernel, Processor and interconnect}  & \bfseries{Compiler and MPI libraries} \\
\hline
Titan				&  US Department of Energy Oak Ridge National Laboratory Leadership Computing Facility & Cray XK7 & 3.0.101-0.47.106.59-default, Opteron 6274 16C 2.200GHz, Cray Gemini interconnect & Intel/17.0.0.098, cray-mpich/7.6.3 \\
\hline
Centennial & US Army Research Laboratory DoD Supercomputing Resource Center & HPE SGI ICE XA & 
	 3.10.0-862.14.4.el7.x86\_64, Xeon E5-2698v4 20C 2.2GHz, Infiniband EDR &
	 Intel/2017.1.132, mpt/2.15 \\
\hline
Excalibur &
US Army Research Laboratory DoD Supercomputing Resource Center &
Cray XC40 &
3.0.101-0.47.106.50-default, Xeon E5-2698v3, Aries &
Intel/17.0.1.132, cray-mpich/7.2.4 \\
\hline

\end{tabularx}
\end{table*}

\section{Conclusions}

The Super Instruction Architecture is a robust, domain-agnostic platform which allows the domain subject matter expert to develop scalable and efficient code while removing the necessity of the domain expert to also be a parallel software developer.  Applications from two very different domains with disparate computational requirements have been recently implemented in the SIA, the quantum chemistry program Aces4 and the T \& D with chemistry program MATLOC. Both codes have been shown to scale well, and are highly competitive with related community codes. A part of this study has shown that the SIA is flexible enough to handle computational complexity, data management flexibility and efficiency as required by the specific domain at hand. Although not reported in this paper, the SIA has also been used to parallelize legacy serial code in a third application domain.  This was easily done by adding wrappers to existing routines to convert them into super instructions, and then writing the necessary SIAL code to orchestrate the parallel computation. 

Aces4, an open source redesign of the long standing community code ACESIII, is reaching the point of being ready for operational use, specially when there is a need for computing open-shell coupled cluster energies. The performance of Aces4 has surpassed its predecessor ACESIII, and the more flexible implementation of the SIA has already enabled new physics to be implemented \cite{aces4IPDPS17}.  

MATLOC is a new implementation of the standard Lagrangian puff transport and dispersion model with the primary design goal of enabling very large climatological studies of mass released into the atmosphere. It is capable of computing millions of releases in just a few hours of wallclock time, enabling finer grain statistics to be generated for confidence bounds of probable mass transport.

\bibliographystyle{IEEEtr}

\begin{mcitethebibliography}{30}
\providecommand*\natexlab[1]{#1}
\providecommand*\mciteSetBstSublistMode[1]{}
\providecommand*\mciteSetBstMaxWidthForm[2]{}
\providecommand*\mciteBstWouldAddEndPuncttrue
  {\def\EndOfBibitem{\unskip.}}
\providecommand*\mciteBstWouldAddEndPunctfalse
  {\let\EndOfBibitem\relax}
\providecommand*\mciteSetBstMidEndSepPunct[3]{}
\providecommand*\mciteSetBstSublistLabelBeginEnd[3]{}
\providecommand*\EndOfBibitem{}
\mciteSetBstSublistMode{f}
\mciteSetBstMaxWidthForm{subitem}{(\alph{mcitesubitemcount})}
\mciteSetBstSublistLabelBeginEnd
  {\mcitemaxwidthsubitemform\space}
  {\relax}
  {\relax}

\bibitem[Lotrich \latin{et~al.}(2005)Lotrich, Ponton, Wang, Yau, Flocke,
  Perera, Deumens, and Bartlett]{lotrich2005}
Lotrich,~V.; Ponton,~M.; Wang,~L.; Yau,~A.; Flocke,~N.; Perera,~A.;
  Deumens,~E.; Bartlett,~R.~J. The super instruction processor parallel design
  pattern for data and floating point intensive algorithms. Workshop on
  Patterns in HPC, University of Illinois at Urbana-Champaign, May 4-6.
  2005\relax
\mciteBstWouldAddEndPuncttrue
\mciteSetBstMidEndSepPunct{\mcitedefaultmidpunct}
{\mcitedefaultendpunct}{\mcitedefaultseppunct}\relax
\EndOfBibitem
\bibitem[Lotrich \latin{et~al.}(2008)Lotrich, Flocke, Ponton, Yau, Perera,
  Deumens, and Bartlett]{lotrich2008chem}
Lotrich,~V.; Flocke,~N.; Ponton,~M.; Yau,~A.~D.; Perera,~A.; Deumens,~E.;
  Bartlett,~R.~J. Parallel Implementation of Electronic Structure Energy,
  Gradient and {Hessian} Calculations. \emph{Journal of Chemical Physics}
  \textbf{2008}, \emph{128}, 194104 (15 pages)\relax
\mciteBstWouldAddEndPuncttrue
\mciteSetBstMidEndSepPunct{\mcitedefaultmidpunct}
{\mcitedefaultendpunct}{\mcitedefaultseppunct}\relax
\EndOfBibitem
\bibitem[Sanders \latin{et~al.}(2010)Sanders, Bartlett, Deumens, Lotrich, and
  Ponton]{Sanders:2010:BLR:1884643.1884677}
Sanders,~B.~A.; Bartlett,~R.; Deumens,~E.; Lotrich,~V.; Ponton,~M. A
  Block-Oriented Language and Runtime System for Tensor Algebra with Very Large
  Arrays. SC '10. 2010\relax
\mciteBstWouldAddEndPuncttrue
\mciteSetBstMidEndSepPunct{\mcitedefaultmidpunct}
{\mcitedefaultendpunct}{\mcitedefaultseppunct}\relax
\EndOfBibitem
\bibitem[Lotrich \latin{et~al.}(2008)Lotrich, Flocke, Ponton, Yau, Perera,
  Deumens, and Bartlett]{acesiii2008}
Lotrich,~V.; Flocke,~N.; Ponton,~M.; Yau,~A.; Perera,~A.; Deumens,~E.;
  Bartlett,~R. Parallel Implementation of Electronic Structure Energy,
  Gradient, and Hessian Calculations. \emph{Journal of Chemical Physics}
  \textbf{2008}, \emph{128}, 2722--2736\relax
\mciteBstWouldAddEndPuncttrue
\mciteSetBstMidEndSepPunct{\mcitedefaultmidpunct}
{\mcitedefaultendpunct}{\mcitedefaultseppunct}\relax
\EndOfBibitem
\bibitem[{Sanders} \latin{et~al.}(2019){Sanders}, {Byrd}, {Jindal}, {Lotrich},
  {Lyakh}, {Perera}, and {Bartlett}]{aces4github}
{Sanders},~B.~A.; {Byrd},~J.~N.; {Jindal},~N.; {Lotrich},~V.~F.; {Lyakh},~D.;
  {Perera},~A.; {Bartlett},~R.~J. Aces4. 2019;
  https://github.com/UFParLab/aces4\relax
\mciteBstWouldAddEndPuncttrue
\mciteSetBstMidEndSepPunct{\mcitedefaultmidpunct}
{\mcitedefaultendpunct}{\mcitedefaultseppunct}\relax
\EndOfBibitem
\bibitem[{Sanders} \latin{et~al.}(2017){Sanders}, {Byrd}, {Jindal}, {Lotrich},
  {Lyakh}, {Perera}, and {Bartlett}]{aces4IPDPS17}
{Sanders},~B.~A.; {Byrd},~J.~N.; {Jindal},~N.; {Lotrich},~V.~F.; {Lyakh},~D.;
  {Perera},~A.; {Bartlett},~R.~J. Aces4: A Platform for Computational Chemistry
  Calculations with Extremely Large Block-Sparse Arrays. 2017 IEEE
  International Parallel and Distributed Processing Symposium (IPDPS). 2017; pp
  555--564\relax
\mciteBstWouldAddEndPuncttrue
\mciteSetBstMidEndSepPunct{\mcitedefaultmidpunct}
{\mcitedefaultendpunct}{\mcitedefaultseppunct}\relax
\EndOfBibitem
\bibitem[Byrd \latin{et~al.}(2018)Byrd, Bartlett, and Sanders]{exascale2018}
Byrd,~J.; Bartlett,~R.; Sanders,~B. In \emph{Exascale Scientific Applications,
  Scalability and Performance Portability}; Straatsma,~T.~P., Antypas,~K.~B.,
  Williams,~T.~J., Eds.; CRC Press: Boca Raton, FL, 2018; Chapter 7, pp
  151--164\relax
\mciteBstWouldAddEndPuncttrue
\mciteSetBstMidEndSepPunct{\mcitedefaultmidpunct}
{\mcitedefaultendpunct}{\mcitedefaultseppunct}\relax
\EndOfBibitem
\bibitem[Chen \latin{et~al.}(2014)Chen, Dinan, Tang, Balaji, Zhong, Wei, Huang,
  and Qin]{Chen:2014:MDM:2683593.2683648}
Chen,~Z.; Dinan,~J.; Tang,~Z.; Balaji,~P.; Zhong,~H.; Wei,~J.; Huang,~T.;
  Qin,~F. MC-checker: Detecting Memory Consistency Errors in MPI One-sided
  Applications. SC '14. 2014; pp 499--510\relax
\mciteBstWouldAddEndPuncttrue
\mciteSetBstMidEndSepPunct{\mcitedefaultmidpunct}
{\mcitedefaultendpunct}{\mcitedefaultseppunct}\relax
\EndOfBibitem
\bibitem[Skamarock \latin{et~al.}(2008)Skamarock, Klemp, Dudhia, Gill, Barker,
  Duda, Huang, Wang, and Powers]{WRF39}
Skamarock,~W.~C.; Klemp,~J.~B.; Dudhia,~J.; Gill,~D.~O.; Barker,~D.~M.;
  Duda,~M.~G.; Huang,~X.-Y.; Wang,~W.; Powers,~J.~G. \emph{A Description of the
  Advanced Research WRF Version 3}; 2008; p 113, DOI: 10.5065/D68S4MVH\relax
\mciteBstWouldAddEndPuncttrue
\mciteSetBstMidEndSepPunct{\mcitedefaultmidpunct}
{\mcitedefaultendpunct}{\mcitedefaultseppunct}\relax
\EndOfBibitem
\bibitem[Harrison(1993)]{harrison1993}
Harrison,~R.~J. Global Arrays. \emph{Theoret. Chim. Acta} \textbf{1993},
  \emph{84}, 363--375\relax
\mciteBstWouldAddEndPuncttrue
\mciteSetBstMidEndSepPunct{\mcitedefaultmidpunct}
{\mcitedefaultendpunct}{\mcitedefaultseppunct}\relax
\EndOfBibitem
\bibitem[Valiev \latin{et~al.}(2010)Valiev, Bylaska, Govind, Kowalski,
  Straatsma, Dam, Wang, Nieplocha, Apra, Windus, and de~Jong]{nwchem2010}
Valiev,~M.; Bylaska,~E.; Govind,~N.; Kowalski,~K.; Straatsma,~T.; Dam,~H.~V.;
  Wang,~D.; Nieplocha,~J.; Apra,~E.; Windus,~T.; de~Jong,~W. NWChem: A
  comprehensive and scalable open-source solution for large scale molecular
  simulations. \emph{Computer Physics Communications} \textbf{2010},
  \emph{181}, 1477 -- 1489\relax
\mciteBstWouldAddEndPuncttrue
\mciteSetBstMidEndSepPunct{\mcitedefaultmidpunct}
{\mcitedefaultendpunct}{\mcitedefaultseppunct}\relax
\EndOfBibitem
\bibitem[Calvin and Valeev(2019)Calvin, and Valeev]{tiledarray}
Calvin,~J.~A.; Valeev,~E.~F. \relax
\mciteBstWouldAddEndPunctfalse
\mciteSetBstMidEndSepPunct{\mcitedefaultmidpunct}
{}{\mcitedefaultseppunct}\relax
\EndOfBibitem
\bibitem[Harrison \latin{et~al.}(2015)Harrison, Beylkin, Bischoff, Calvin,
  Fann, Fosso{-}Tande, Galindo, Hammond, Hartman{-}Baker, Hill, Jia, Kottmann,
  Ou, Ratcliff, Reuter, Richie{-}Halford, Romero, Sekino, Shelton, Sundahl,
  Thornton, Valeev, V{\'{a}}zquez{-}Mayagoitia, Vence, and
  Yokoi]{DBLP:journals/corr/HarrisonBBCFFGH15}
Harrison,~R.~J. \latin{et~al.}  {MADNESS:} {A} Multiresolution, Adaptive
  Numerical Environment for Scientific Simulation. \emph{CoRR} \textbf{2015},
  \emph{abs/1507.01888}\relax
\mciteBstWouldAddEndPuncttrue
\mciteSetBstMidEndSepPunct{\mcitedefaultmidpunct}
{\mcitedefaultendpunct}{\mcitedefaultseppunct}\relax
\EndOfBibitem
\bibitem[Peng \latin{et~al.}(2019)Peng, Calvin, and Valeev]{peng2019}
Peng,~C.; Calvin,~J.~A.; Valeev,~E.~F. Coupled-cluster singles, doubles and
  perturbative triples with density fitting approximation for massively
  parallel heterogeneous platforms. \emph{International Journal of Quantum
  Chemistry} \textbf{2019}, e25894\relax
\mciteBstWouldAddEndPuncttrue
\mciteSetBstMidEndSepPunct{\mcitedefaultmidpunct}
{\mcitedefaultendpunct}{\mcitedefaultseppunct}\relax
\EndOfBibitem
\bibitem[Bartlett and Musia\l(2007)Bartlett, and Musia\l]{bartlett2007}
Bartlett,~R.~J.; Musia\l,~M. {Coupled-Cluster Theory in Quantum Chemistry}.
  \emph{Reviews in Modern Physics} \textbf{2007}, \emph{79}, 291--352\relax
\mciteBstWouldAddEndPuncttrue
\mciteSetBstMidEndSepPunct{\mcitedefaultmidpunct}
{\mcitedefaultendpunct}{\mcitedefaultseppunct}\relax
\EndOfBibitem
\bibitem[Shavitt and Bartlett(2009)Shavitt, and Bartlett]{shavitt2009}
Shavitt,~I.; Bartlett,~R.~J. \emph{Many-Body Methods in Chemistry and Physics};
  Cambridge: New York, 2009\relax
\mciteBstWouldAddEndPuncttrue
\mciteSetBstMidEndSepPunct{\mcitedefaultmidpunct}
{\mcitedefaultendpunct}{\mcitedefaultseppunct}\relax
\EndOfBibitem
\bibitem[Purvis and Bartlett(1982)Purvis, and Bartlett]{purvis1982}
Purvis,~G.; Bartlett,~R. A full coupled-cluster-singles and doubles model: The
  inclusion of disconnected triples. \emph{J. Chem. Phys.} \textbf{1982},
  \emph{76}, 1910--1918\relax
\mciteBstWouldAddEndPuncttrue
\mciteSetBstMidEndSepPunct{\mcitedefaultmidpunct}
{\mcitedefaultendpunct}{\mcitedefaultseppunct}\relax
\EndOfBibitem
\bibitem[K.~Ragavachari and Head-Gordon(1989)K.~Ragavachari, and
  Head-Gordon]{Ragavachari}
K.~Ragavachari,~J. A.~P.,~G. W.~Trucks; Head-Gordon,~M. A fifth-order
  perturbation comparison of electron correlation theories. \emph{Chemical
  Physics Letters} \textbf{1989}, \emph{157}, 479\relax
\mciteBstWouldAddEndPuncttrue
\mciteSetBstMidEndSepPunct{\mcitedefaultmidpunct}
{\mcitedefaultendpunct}{\mcitedefaultseppunct}\relax
\EndOfBibitem
\bibitem[Anisimov \latin{et~al.}(2014)Anisimov, Bauer, Chadalavada, Olson,
  Glenski, Kramer, Apr\'{a}, and Kowalski]{anisimov2014}
Anisimov,~V.~M.; Bauer,~G.~H.; Chadalavada,~K.; Olson,~R.~M.; Glenski,~J.~W.;
  Kramer,~W. T.~C.; Apr\'{a},~E.; Kowalski,~K. Optimization of the Coupled
  Cluster Implementation in NWChem on Petascale Parallel Architectures.
  \emph{Journal of Chemical Theory and Computation} \textbf{2014}, \emph{10},
  4307--4316\relax
\mciteBstWouldAddEndPuncttrue
\mciteSetBstMidEndSepPunct{\mcitedefaultmidpunct}
{\mcitedefaultendpunct}{\mcitedefaultseppunct}\relax
\EndOfBibitem
\bibitem[Peng \latin{et~al.}(2019)Peng, Lewis, Wang, Clement, Pavosevic, Zhang,
  Rishi, Teke, Pierce, Calvin, Kenny, Seidl, Janssen, and Valeev]{MPQC}
Peng,~C.; Lewis,~C.; Wang,~X.; Clement,~M.; Pavosevic,~F.; Zhang,~J.;
  Rishi,~V.; Teke,~N.; Pierce,~K.; Calvin,~J.; Kenny,~J.; Seidl,~E.;
  Janssen,~C.; Valeev,~E. \relax
\mciteBstWouldAddEndPunctfalse
\mciteSetBstMidEndSepPunct{\mcitedefaultmidpunct}
{}{\mcitedefaultseppunct}\relax
\EndOfBibitem
\bibitem[Seely \latin{et~al.}(2008)Seely, Kienzle, and Masters]{SLAM2008}
Seely,~S.; Kienzle,~M.; Masters,~S. \emph{Technical Documentation for the Short
  Range Layered Atmospheric Model SLAM Version 4.20}; 2008; p 156, Available
  from ENSCO, 4849 Wickham Rd., Melbourne, FL 32940\relax
\mciteBstWouldAddEndPuncttrue
\mciteSetBstMidEndSepPunct{\mcitedefaultmidpunct}
{\mcitedefaultendpunct}{\mcitedefaultseppunct}\relax
\EndOfBibitem
\bibitem[Holmes and Morawska(2006)Holmes, and Morawska]{HOLMES20065902}
Holmes,~N.; Morawska,~L. A review of dispersion modeling and its application to
  the dispersion of particles: An overview of different dispersion models
  available. \emph{Atmospheric Environment} \textbf{2006}, \emph{40}, 5902 --
  5928\relax
\mciteBstWouldAddEndPuncttrue
\mciteSetBstMidEndSepPunct{\mcitedefaultmidpunct}
{\mcitedefaultendpunct}{\mcitedefaultseppunct}\relax
\EndOfBibitem
\bibitem[Sykes \latin{et~al.}(1999)Sykes, Cerasoli, and Henn]{scipuff}
Sykes,~R.~I.; Cerasoli,~C.~P.; Henn,~D.~S. The representation of dynamic flow
  effects in a Lagrangian puff dispersion model. \emph{Journal of Hazardous
  Materials} \textbf{1999}, \emph{64}, 223--247\relax
\mciteBstWouldAddEndPuncttrue
\mciteSetBstMidEndSepPunct{\mcitedefaultmidpunct}
{\mcitedefaultendpunct}{\mcitedefaultseppunct}\relax
\EndOfBibitem
\bibitem[Stein \latin{et~al.}(2015)Stein, Draxler, Rolph, Stunder, Cohen, and
  Ngan]{stein2015}
Stein,~A.~F.; Draxler,~R.~R.; Rolph,~G.~D.; Stunder,~B. J.~B.; Cohen,~M.~D.;
  Ngan,~F. NOAA's HYSPLIT Atmospheric Transport and Dispersion Modeling System.
  \emph{Bulletin of the American Meteorological Society} \textbf{2015},
  \emph{96}, 2059--2077\relax
\mciteBstWouldAddEndPuncttrue
\mciteSetBstMidEndSepPunct{\mcitedefaultmidpunct}
{\mcitedefaultendpunct}{\mcitedefaultseppunct}\relax
\EndOfBibitem
\bibitem[Rolph \latin{et~al.}(2017)Rolph, Stein, and Stunder]{rolph2017}
Rolph,~G.; Stein,~A.; Stunder,~B. Real-time Environmental Applications and
  Display sYstem: READY. \emph{Environmental Modeling \& Software}
  \textbf{2017}, \emph{95}, 210--228\relax
\mciteBstWouldAddEndPuncttrue
\mciteSetBstMidEndSepPunct{\mcitedefaultmidpunct}
{\mcitedefaultendpunct}{\mcitedefaultseppunct}\relax
\EndOfBibitem
\bibitem[Pasquill(1971)]{pasquill1971}
Pasquill,~F. Atmospheric dispersion of pollution. \emph{Quarterly Journal of
  the Royal Meteorological Society} \textbf{1971}, \emph{97}, 369--395\relax
\mciteBstWouldAddEndPuncttrue
\mciteSetBstMidEndSepPunct{\mcitedefaultmidpunct}
{\mcitedefaultendpunct}{\mcitedefaultseppunct}\relax
\EndOfBibitem
\bibitem[Dunning~Jr(1989)]{dunning1989}
Dunning~Jr,~T. Gaussian Basis Sets for Use in Correlated Molecular
  Calculations. I. The Atoms Boron Through Neon and Hydrogen. \emph{J. Chem.
  Phys.} \textbf{1989}, \emph{90}, 1007--1023\relax
\mciteBstWouldAddEndPuncttrue
\mciteSetBstMidEndSepPunct{\mcitedefaultmidpunct}
{\mcitedefaultendpunct}{\mcitedefaultseppunct}\relax
\EndOfBibitem
\bibitem[Krishnan \latin{et~al.}(1980)Krishnan, Binkley, Seeger, and
  Pople]{krishnan1980}
Krishnan,~R.; Binkley,~J.~S.; Seeger,~R.; Pople,~J.~A. Self-consistent
  molecular orbital methods. XX. A basis set for correlated wave functions.
  \emph{The Journal of Chemical Physics} \textbf{1980}, \emph{72},
  650--654\relax
\mciteBstWouldAddEndPuncttrue
\mciteSetBstMidEndSepPunct{\mcitedefaultmidpunct}
{\mcitedefaultendpunct}{\mcitedefaultseppunct}\relax
\EndOfBibitem
\bibitem[McLean and Chandler(1980)McLean, and Chandler]{mcLean1980}
McLean,~A.~D.; Chandler,~G.~S. Contracted Gaussian basis sets for molecular
  calculations. I. Second row atoms, Z=11–18. \emph{The Journal of Chemical
  Physics} \textbf{1980}, \emph{72}, 5639--5648\relax
\mciteBstWouldAddEndPuncttrue
\mciteSetBstMidEndSepPunct{\mcitedefaultmidpunct}
{\mcitedefaultendpunct}{\mcitedefaultseppunct}\relax
\EndOfBibitem
\end{mcitethebibliography}
\providecommand{\latin}[1]{#1}
\makeatletter
\providecommand{\doi}
  {\begingroup\let\do\@makeother\dospecials
  \catcode`\{=1 \catcode`\}=2\doi@aux}
\providecommand{\doi@aux}[1]{\endgroup\texttt{#1}}
\makeatother
\providecommand*\mcitethebibliography{\thebibliography}
\csname @ifundefined\endcsname{endmcitethebibliography}
  {\let\endmcitethebibliography\endthebibliography}{}


\end{document}